\documentclass{aa}
\usepackage{times,graphics,psfig}

\def\ma{~\raise.5ex\hbox{$>$}\kern-.8em\lower 1mm\hbox{$\sim$}~}
\def\ergs{erg cm$^{\rm -2}$ s$^{\rm -1}$}
\def\iras{IRAS 13349+2438}

\def\nh{$N_{\rm H}$ }
\def\oxa{O\,{\sc viii}}
\def\oxs{O\,{\sc vii}}
\begin{document}
\thesaurus{03(11.01.2; 11.17.4 IRAS 13349+2438; 13.25.2)}
\title{The warm absorber in IRAS 13349+2438: dusty or not?}
\author{J.\,Siebert, S.\,Komossa and W.\,Brinkmann}
\institute{Max-Planck-Institut f\"ur extraterrestrische Physik,
Giessenbachstrasse, D-85740 Garching, Germany}
\mail{J. Siebert, jos@mpe.mpg.de}
\titlerunning{The warm absorber in IRAS 13349+2438}
\authorrunning{Siebert et al.}
\date{Received 28 June 1999; accepted}
\maketitle
\begin{abstract}
\object{\iras} was the first quasar suggested to host a warm absorber with 
internal dust. We obtained new HRI data for this object and derive constraints on the 
X-ray emitting region of \iras\ by investigating its X-ray extent and lightcurve. Rapid
X-ray variability is detected with a factor of two change in count rate within one 
day and 20\% variability within about three hours, which practically rules out any significant
contribution of scattered X-rays. We present for the first time a detailed, self-consistent 
modeling of the ROSAT PSPC and ASCA spectrum of \iras\ in terms of a warm absorber including 
dust. It turns out that both a dust-free and a dusty single-component warm absorber fail 
to explain the ROSAT, ASCA, and optical data simultaneously. We discuss possible explanations 
such as a variable and/or complex warm absorber, a variable soft X-ray continuum and instrumental 
effects. 
\end{abstract}


\section{Introduction}

\iras\ is the archetypal infrared-selected quasar, discovered in a sample of bright 
IRAS 12$\mu$ sources by Beichmann et al. (\cite{beichmann}). It is optically bright 
($m_{\rm V} = 14.7$) and located at a redshift of $z = 0.10764\pm 0.00027$ (Kim et al.
\cite{kim}). Broadband investigations revealed that the spectral energy distribution
is dominated by infrared emission. Its bolometric luminosity\footnote{$H_{\rm 0} =
50$ km s$^{-1}$ Mpc$^{-1}$ and $q_{\rm 0} = 0.5$ are assumed throughout this paper.}
between 0.34 and 100$\mu$ is $\approx 2.3\times 10^{46}$ erg s$^{-1}$ (Beichmann et 
al. \cite{beichmann}). VLA observations show a weak, unresolved radio source with a 
flux density of $\sim 6$ mJy at 4.85 GHz (Laurent-Muehleisen et al. \cite{sally}). 

Wills et al. (\cite{wills}) presented polarimetric and spectrophotometric 
observations and found high polarization which rises from 1.4\% in the K-band 
(2.2$\mu$) to 8\% in the U-band (0.36$\mu$). From the very large hydrogen broad-line 
ratios ($H_{\alpha}/H_{\beta} = 5.9$; $Pa_{\alpha}/H_{\alpha} = 0.15$) and the shape 
of the optical continuum they derive $E(B-V)\approx 0.3$, which 
corresponds to a hydrogen column density of $1.7\times 10^{21}$ cm$^{-2}$ assuming 
standard dust and a Galactic gas-to-dust ratio. To explain their observations, Wills 
et al. (\cite{wills}) argued for a model which invokes a bipolar geometry of the 
central region and in which the line-of-sight just grazes the edge of a dusty, 
parsec-scale torus or disk. The observed optical emission is a superposition 
of two components: a direct one, which gets attenuated as the light passes through the 
molecular torus, and a second one, which is emitted in polar directions and then 
scattered towards us by electrons or small dust grains. The latter component is 
therefore polarized parallel to the major axis of the host galaxy.

\iras\ turned out to be an extraordinarily bright and variable soft X-ray source. 
In the ROSAT All-Sky Survey it was detected with $\approx 2.5$ cts/s in the 0.1-2.4 
keV energy band (Brinkmann \& Siebert \cite{wpb94}). Brandt et al.
(\cite{brandt96}; BFP96 hereafter) analyzed two short ROSAT PSPC pointed observations and 
found large amplitude variability as well as evidence for an ionized absorber in the 
soft X-ray spectrum by identifying the \oxa\ absorption edge at (rest frame) 
0.871 keV. They also suggested the presence of dust within the so-called 'warm absorber' 
in order to explain the apparent discrepancy between the observed optical reddening 
and the absence of any cold absorption in excess of the Galactic \nh value.

Whereas the model of a warm absorber with internal dust has meanwhile been 
successfully applied to the X-ray spectra of several AGN (e.g. NGC 3227, NGC 3786, 
IRAS 17020+4544; Komossa \& Fink 1997a,b; Leighly et al.
\cite{leighly97b}; Komossa \& Bade \cite{komossa98}) this has not yet been done for 
\iras (for first results see Komossa \& Greiner \cite{komossa99}). Given the potentially 
strong modifications of the soft X-ray spectrum by the 
presence of dust (e.g. Komossa \& Fink \cite{komossa97a}) it is important to 
scrutinize whether (and under which conditions) a dusty warm absorber is consistent 
with the observed X-ray spectrum. ROSAT spectra are best suited to accomplish this, 
since the most pronounced signature of dust is the carbon edge at 0.28 keV, which is 
outside the ASCA energy range.

Brinkmann et al. (\cite{wpb96}) analyzed the ASCA observation of \iras\ and found 
distinct changes in the spectral parameters compared to the earlier ROSAT data. These
findings were confirmed by Brandt et al. (\cite{brandt97}) in an independent analysis
of the data. In particular, the strong \oxa\ absorption edge found in the 
PSPC observation was not detected anymore. Instead, both analyses find a spectral 
feature around $\sim 0.65$ keV, which is interpreted as an emission line by Brinkmann
et al. (\cite{wpb96}) and a weak \oxs\ absorption edge by Brandt et al. 
(\cite{brandt97}). 

Besides a successful spectral fit, a key argument for invoking the presence of {\em 
dusty} warm absorbers is the apparent discrepancy between the column densities of 
cold material derived from the optical reddening and the absence of excess soft X-ray
absorption. A crucial assumption for this argument is that both, the optical and 
the X-ray continua, are seen along the same line of sight. The necessity
of {\em dusty} warm material would be alleviated if the X-ray emission were spatially 
extended, i.e. if the X-rays would not pass through the material causing the optical
extinction. Observations with the ROSAT HRI allow to set limits on the contribution 
of an extended component to the X-ray emission by its improved spatial resolution
and the confirmation of the rapid X-ray variability observed with ASCA (Brinkmann et al.
\cite{wpb96}).

\begin{figure}
\psfig{figure=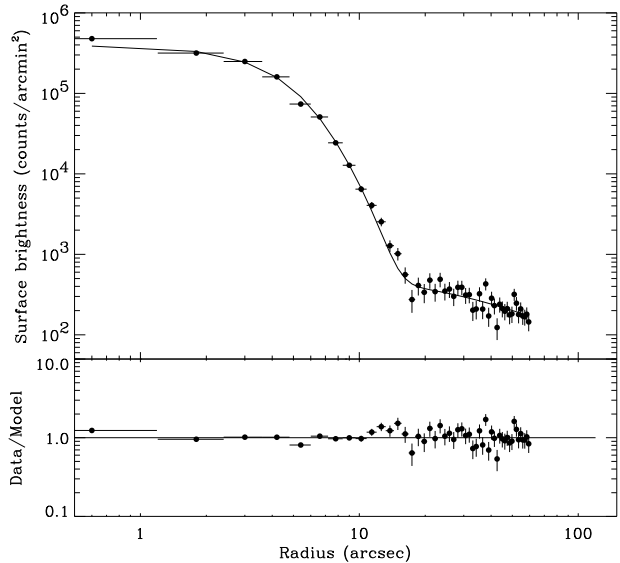,width=9cm}
\caption{Upper panel: the surface brightness profile (filled circles) and the best-fitting
PSF model for IRAS 13349+2438 (solid line). Lower panel: ratio of surface brightness data 
to PSF model.}
\label{psf}
\end{figure}

We note that \iras\ was also considered as a narrow-line Seyfert 1 galaxy (Osterbrock \& Pogge 
\cite{osterbrock}), due to some of its optical (permitted emission line widths, 
[O\,{\footnotesize III}]/H$\beta$ ratio) and X-ray properties (steep spectrum, rapid 
variability), which are typical for this class of AGN. We will discuss some of our results 
in view of this hypothesis.

\section{The ROSAT HRI data}

\iras\ was observed twice with the ROSAT HRI. The first observation (hereafter H1) 
took place in 1996 from July 16 to July 21 and resulted in an effective exposure of 
6209 s. The second observation (H2) was conducted roughly one year later, from June 
15 to June 24 1997, and resulted in 24748 s of good exposure. \iras\ was observed 
on-axis in both observations with an average count rate of $\approx 0.3$ cts/s.

\subsection{Spatial extent}

The high spatial resolution of the ROSAT HRI provides tight constraints on any 
extended emission component in \iras. The HRI point spread function (PSF) can be 
approximated by two Gaussians with $\sigma_{\rm 1}\approx 2.2$ arcsec and 
$\sigma_{\rm 2}\approx 4.0$ arcsec and an exponential term, which is relevant for 
large radii ($\approx 30$ arcsec) and bright sources (David et al. \cite{david}).
The inner core of the PSF corresponds to a projected linear size of $\approx 5.7$ kpc 
at the redshift of \iras. Due to random errors in the aspect solution, the values of 
$\sigma_{\rm 1}$ and $\sigma_{\rm 2}$ vary from 1.9 to 2.5 arcsec and from 3.3 to 4.1 arcsec,
respectively, between individual observations of point sources. Further, elongations of 
the PSF by uncorrected residual wobble motion have been reported for a number of 
sources (Morse \cite{morse}; W.Pietsch, priv. com.).

\begin{figure}
\psfig{figure=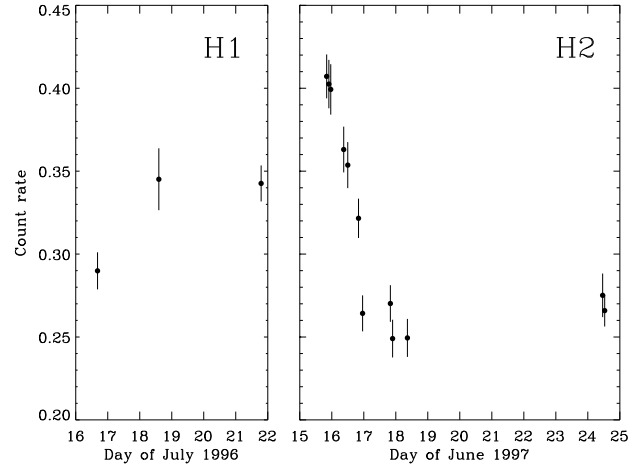,width=8.5cm}
\caption{The HRI light curve of \iras\ for both observations. \iras\ displays 
large amplitude count rate variations on time scales
of hours to days. The fastest variability detected is a 20\% decline in count
rate within $\approx 3$ hours.}
\label{light}
\end{figure}

\begin{figure*}
\psfig{figure=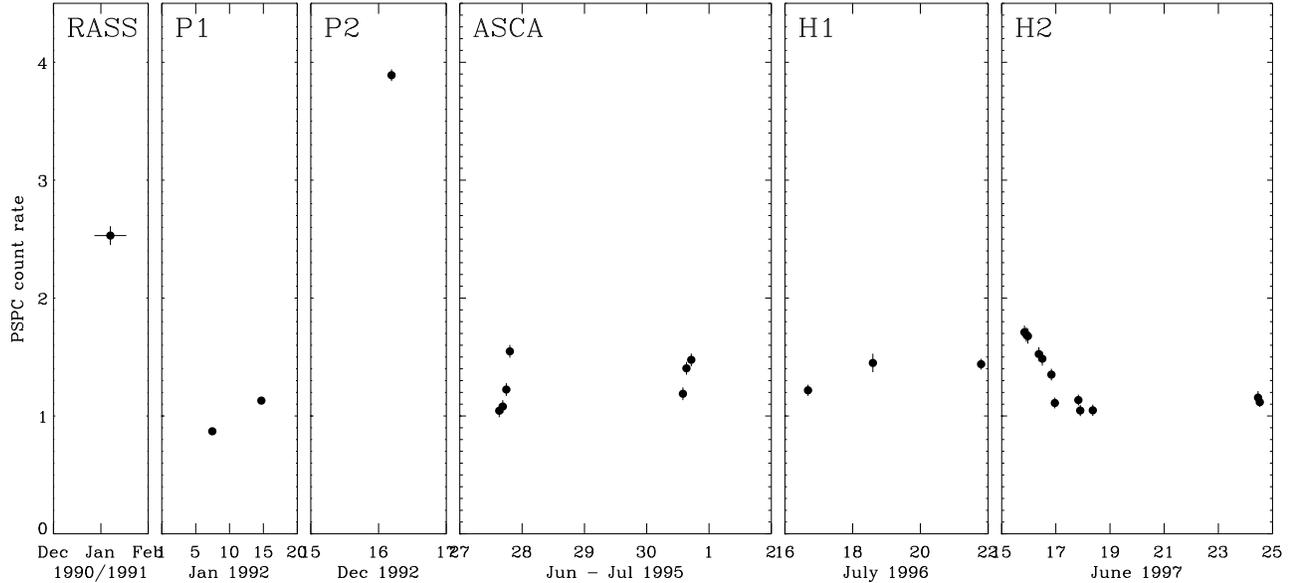,width=18cm}
\caption{The historic X-ray light curve for \iras. Shown are the All-Sky Survey
observation (RASS), both PSPC pointings (P1,P2), the ASCA data and the two HRI
observations (H1,H2). All measurements were converted to ROSAT PSPC count rates
assuming a simple power law spectrum with $\Gamma = 2.8$ and Galactic absorption.}
\label{historic}
\end{figure*}

In a practical approach, the observation dependent uncertainties were modeled by smearing 
the theoretical PSF with an additional Gaussian characterized by $\sigma_+$. We searched 
for the best-fitting $\sigma_+$ by fitting the smeared PSF to the binned radial profile 
of \iras\ from the H2 observation using a least-squares method. The best description, shown
in Fig.~\ref{psf}, is achieved for $\sigma_+\approx 1.8$ arcsec. This value is well within
the range determined from numerous other analyses of point source observations. Given the mentioned 
uncertainties in the determination of the correct HRI PSF for an individual observation, 
we conclude that the X-ray emission of \iras\ is point like to the limit of the HRI 
resolution.

\subsection{Light curves}

Source counts were extracted from a circular region with radius 1.5 arcmin around the
centroid position of the X-ray source. The background was determined from a source 
free annulus with inner radius 3 and outer radius 5 arcmin. Since \iras\ was observed
on-axis, no vignetting correction was applied to the data. Finally, the observations 
were split into individual observation intervals with durations between 1000 to 3000 
s. The intervals were chosen by hand to ensure sufficient photon statistics in each 
data point and to match as closely as possible the distribution of the observation 
intervals in ``real time''. The resulting light curves are shown in Fig.~\ref{light}.
 
The average count rate in both, H1 and H2, is $\approx 0.3$ cts/s. Assuming a simple
power law spectrum with a photon index of $\Gamma = 2.8$ (see Sect. 3.3.1) and Galactic 
absorption (\nh $= (1.1\pm 0.2)\times 10^{20}$ cm$^{-2}$; Murphy et al. \cite{murphy}), this 
count rate yields an unabsorbed 0.1-2.4 keV flux of $8.2\times 10^{-12}$ \ergs. The 
corresponding rest frame 0.1-2.4 keV luminosity is $4.6\times 10^{44}$ erg s$^{-1}$. 

\iras\ clearly displays rapid and large amplitude variability in both HRI 
observations. At the beginning of H2 we obviously observed the decline of a larger 
outburst and the count rate decreases by almost a factor of two within one day. The 
fastest variability seen is a decline by 20\% within about 3 hours, which is truly 
remarkable given the high X-ray luminosity of \iras. Following Lawrence \& Papadakis 
(\cite{lawrence}), who find a correlation between luminosity and doubling time scale 
from EXOSAT variability power spectrum analysis, a variability time scale of the 
order of 10$^{\rm 6}$ to 10$^{\rm 7}$ sec would have been expected for \iras.

For illustrative purposes we show in Fig.~\ref{historic} the historical X-ray
light curve for \iras\ including all available data from the ROSAT All-Sky 
Survey to the latest HRI observations. All measurements have been converted 
to PSPC count rates assuming a simple power law spectrum ($\Gamma = 2.8$) and 
Galactic absorption.

X-ray variability was detected in all previous observations of \iras. BFP96
note a factor of $\approx 4$ change between the two PSPC observations
separated by about one year (P1 and P2). During P2 in December 1992, \iras\ obviously
happened to be in a historical high state. Unfortunately, P2 only lasted $\approx 
1570$ s and no variability was detected within this time interval (BFP96).
In the ASCA observation a 50\% increase in flux within about 5 
hours was detected (Brinkmann et al. \cite{wpb96}). This time scale is similar to 
that found in the HRI observation. 

Using the observed variability time scales and causality arguments we can derive
upper limits on the size of the X-ray emitting region ($R\le c\cdot\Delta t$; 
neglecting relativistic beaming). A time scale of one day gives an upper limit of
$\approx 180$ AU. The fastest observed variability time scales of 3 to 5 hours
suggest a size of the X-ray source not larger than 20 to 30 AU. This 
practically rules out a significant contribution of scattered X-rays, since
the observed sizes of electron scattering mirrors are a factor of $10^4 - 10^5$
larger ($\sim 50$ pc in NGC 1068; Antonucci et al. \cite{antonucci}). In addition, 
a small mirror would imply very high electron densities and a Thomson-thick 
scattering medium, which might be in contradiction to observations at other 
wavelengths (see discussion in BFP96).

Variability also constrains the efficiency of the conversion of infalling matter
into radiation energy (e.g. Fabian \cite{fabian}):
\[
\eta \ga \frac{\Delta L\,[erg/s]}{2.1\times 10^{42}}\; \Delta t[s]^{-1}
\]
The observed change in luminosity of $\Delta L \approx 2.3\times 10^{44}$ erg 
s$^{-1}$ within $\approx 10^4$ s gives $\eta \ga$ 1.2\%.    

We finally note, that the above arguments hold only in the absence of relativistic
beaming, which would effectively shorten the observed variability time scales by time 
dilation effects. Models involving relativistic motion have recently been
proposed for a number of narrow-line Seyfert 1 (NLS1) galaxies to explain spectral 
features (Leighly et al. \cite{leighly97a}; Hayashida \cite{hayashida}) and 
in particular to explain the rapid X-ray variability in IRAS 13224-3809 (Boller et al. 
\cite{boller}). Despite its high luminosity, \iras\ shows many similarities to NLS1 
galaxies, e.g. a steep soft X-ray spectrum and rapid variability. It also satisfies 
the NLS1 classification criteria of Goodrich (\cite{goodrich89b}). Therefore 
the presence of relativistic effects in \iras\ cannot be excluded.  

\section{The ROSAT PSPC data: Spectral analysis}

\subsection{Data preparation}

\iras\ was observed in two short pointings with the ROSAT PSPC from January 7 to
January 14, 1992 ($t_{\rm eff} \approx 3280$ s) and on December 16 1992 ($t_{\rm eff} \approx 
1570$ s). Hereafter the two observations are referred to as P1 and P2, respectively. The 
data were extracted from the ROSAT archive at MPE
and analyzed using standard routines within the EXSAS environment (Zimmermann et al. 
\cite{zimmermann}). Source photons were extracted from a circular cell with 
radius 3 arcmin. The background was determined from a circular source-free region 
near the target source. Corrections for vignetting and dead-time were applied. For the
spectral analysis, the pulse height spectrum between channels 11 and 240 was rebinned
such that the signal-to-noise ratio in each bin was at least 10. 

In the ROSAT All-Sky Survey \iras\ was observed for about $\approx$365 s. We extracted
source photons from a circle with radius 10 arcmin centered on \iras. To ensure a 
similar exposure, the background was determined from a source free circle with the 
same radius, but displaced from \iras\ along the scanning direction of the satellite 
during the All-Sky Survey. Again the data were corrected for vignetting and dead time.
In total, about 930 photons were accumulated for \iras\ during the Survey observation.
Finally, the pulse height spectrum between channels 11 and 240 was rebinned according
to a constant signal-to-noise ratio of 5.h1626.tex

\subsection{Warm absorber models}

\begin{figure}
\psfig{figure=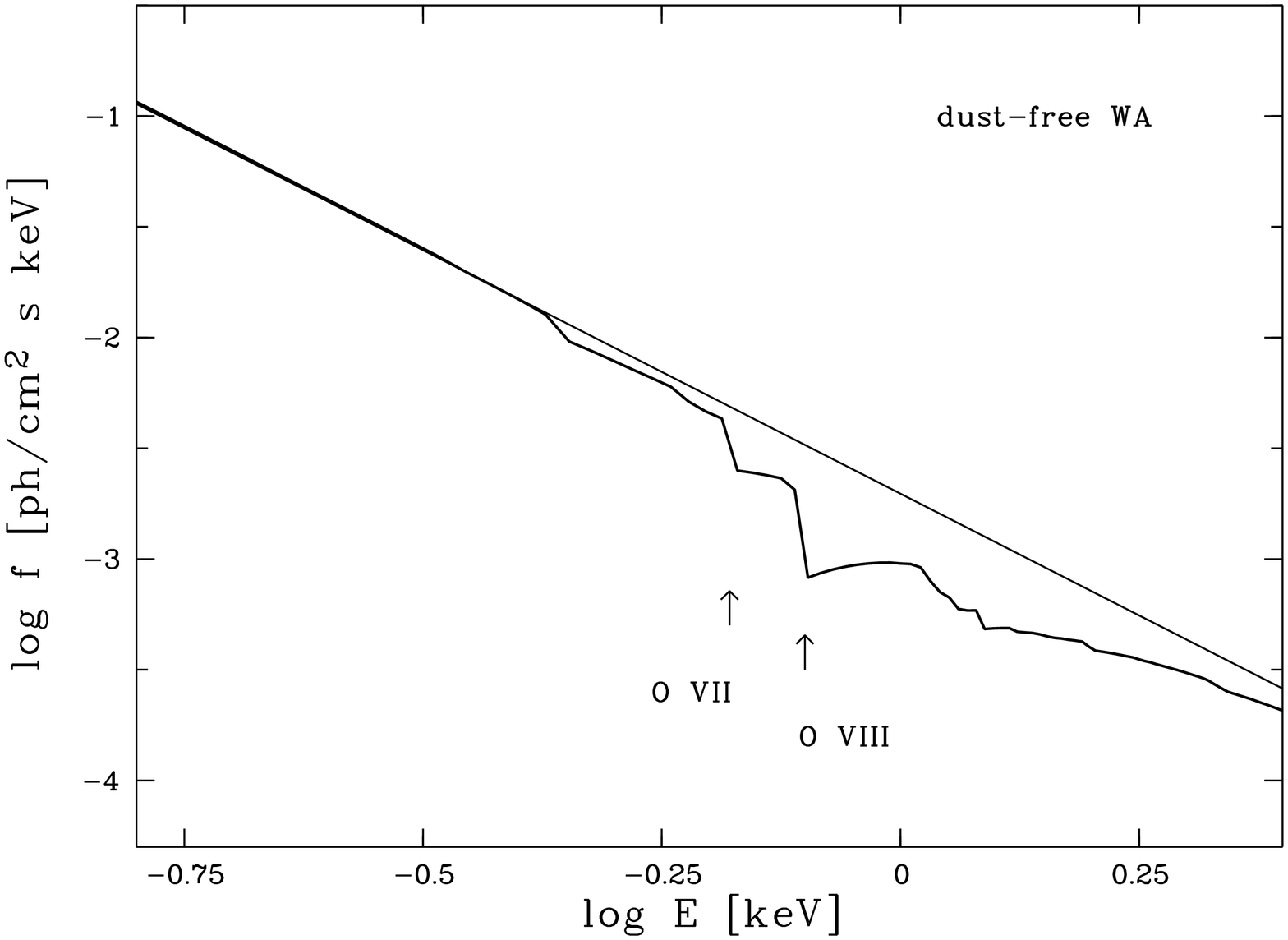,width=8cm,angle=-90}
\psfig{figure=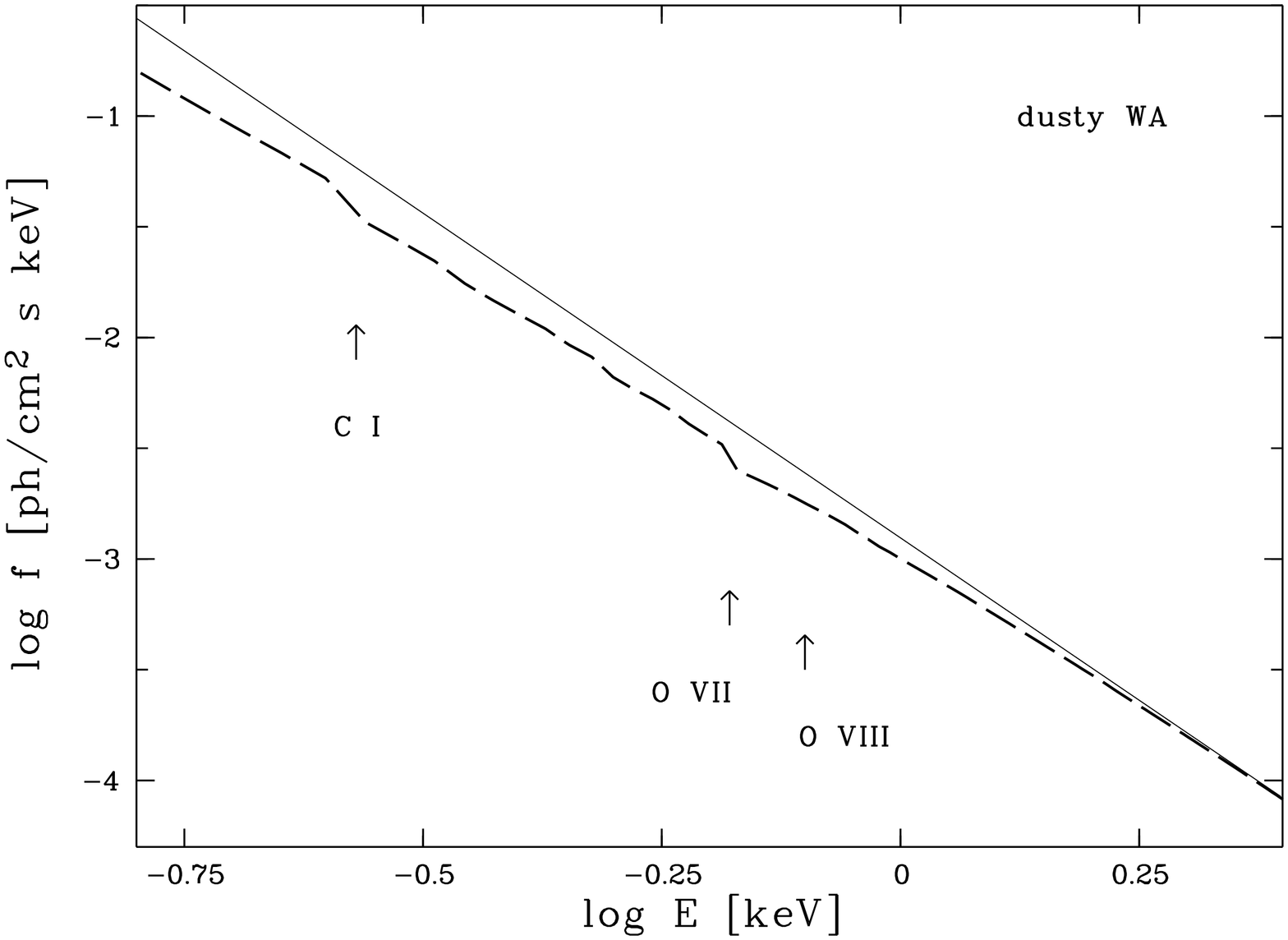,width=8cm,angle=-90}
\caption{Two representative models of a dust-free (top panel) and a dusty (bottom panel)
warm absorber to illustrate the influence of dust. The range of the abscissa corresponds 
to the ROSAT energy band (0.1-2.4 keV). The respective model parameters correspond to 
the best-fit values determined for the ROSAT PSPC observations. In the case of the 
dust-free warm absorber these are $\Gamma = 2.2$ and log N$_{\rm w} = 22.5$. The dusty 
warm absorber model is shown for log N$_{\rm w}$ = log N$_{\rm opt}$ = 21.2 and $\Gamma = 2.9$. 
All model spectra have been corrected for Galactic cold absorption. The straight lines mark 
the respective {\em intrinsic} power law spectra. See text for further details.}
\label{wa-dwa}
\end{figure}

To discuss the X-ray spectrum in terms of warm absorption we used warm absorber 
models calculated with the photoionization code {\sc Cloudy} (Ferland \cite{ferland}). 
For the general 
assumptions of the models see Komossa \& Fink (1997a,b). 
Briefly, the ionized material was assumed to be of constant density (log $n_{\rm H}$ 
= 7) and of solar abundances (Grevesse \& Anders \cite{grevesse}) in the dust-free case. 
A typical Seyfert IR to $\gamma$-ray spectral energy distribution (with $\alpha_{\rm uv-x}=
-1.4$ in the EUV) was assumed as ionizing continuum. In the warm absorber models including dust, 
the dust composition and grain size distribution were like those of the Galactic diffuse 
interstellar medium (Mathis et al. \cite{mathis}), and the chemical abundances 
were depleted correspondingly, if not mentioned otherwise. We focused on Galactic-ISM-like
dust, because this was also assumed in the optical reddening estimates (e.g. Brandt et al.
1996). To test specific scenarios we also varied the dust properties, but these are not treated 
as additional free fit parameters. 
The warm absorber properties which are 
derived from X-ray spectral fits are the ionization parameter $U=Q/(4\pi{r}^{2}n_{\rm H}c)$
and the hydrogen column density $N_{\rm w}$ of the ionized material, where $Q$ is the count 
rate of incident photons above the Lyman limit, $r$ is the distance between the central 
source and the warm absorber and $n_{\rm H}$ is the hydrogen density. 

\begin{figure*}
\begin{minipage}{9cm}
\psfig{figure=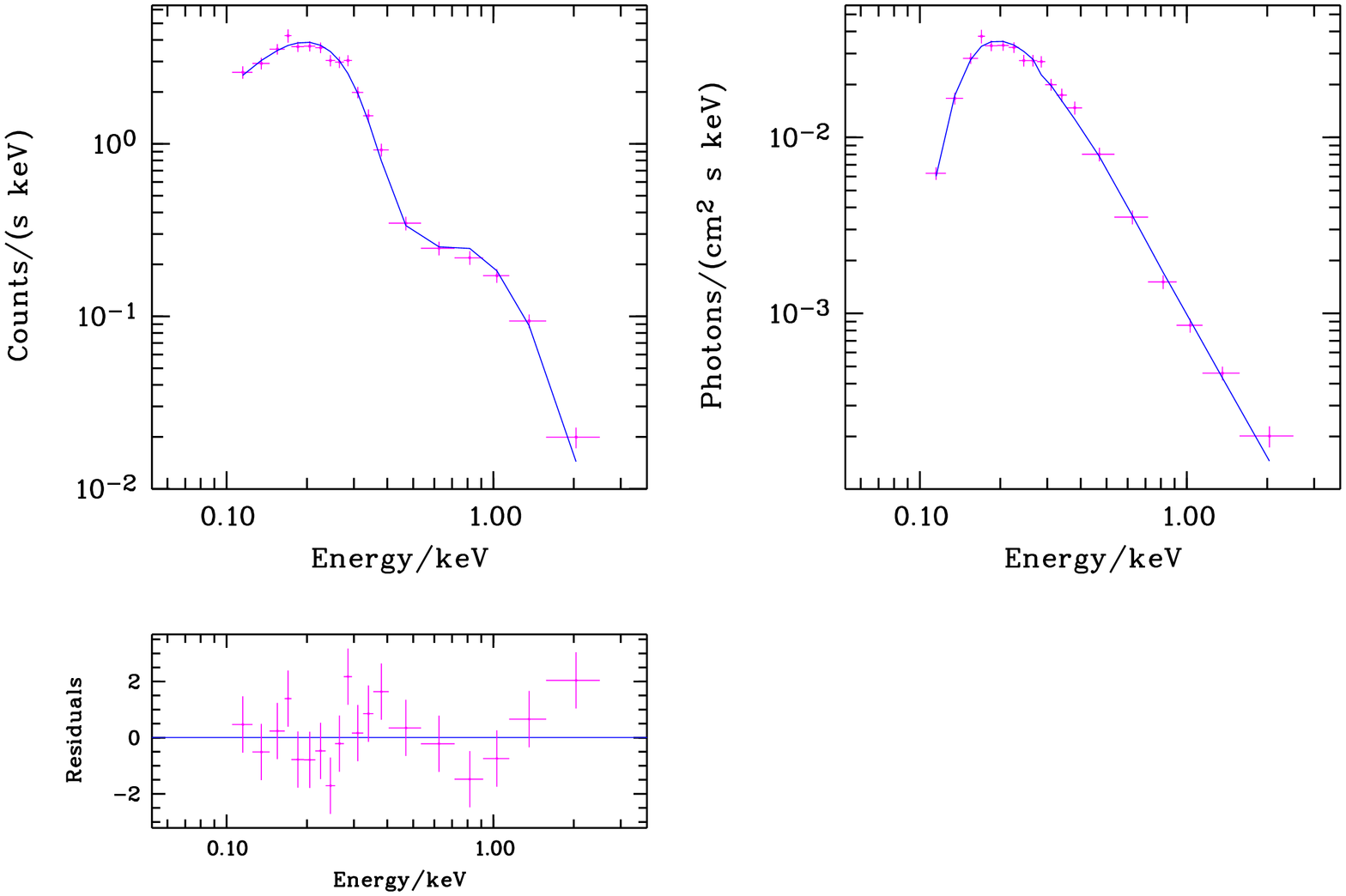,angle=-90,width=9cm,clip=}
\psfig{figure=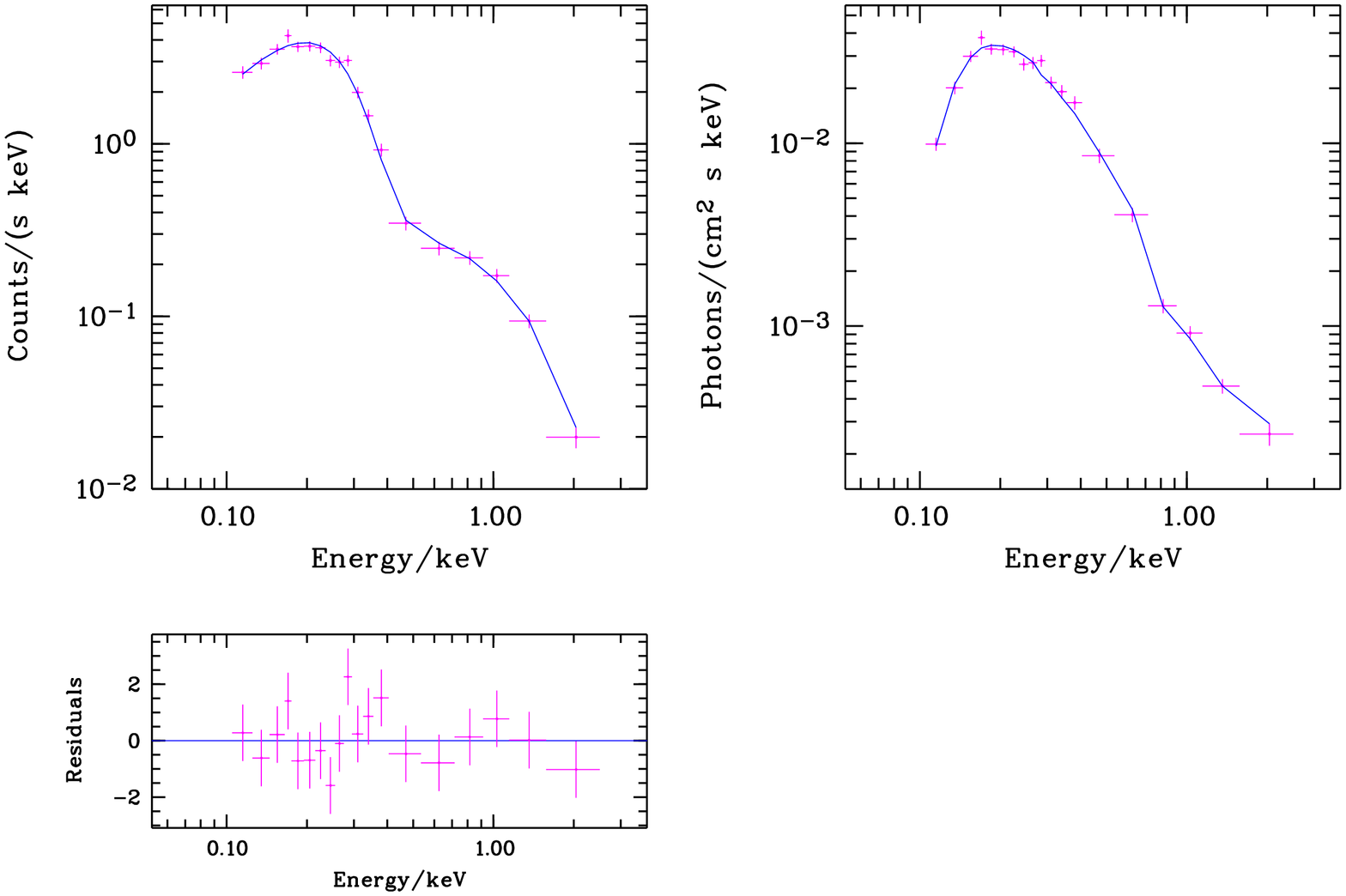,angle=-90,width=9cm,clip=}
\psfig{figure=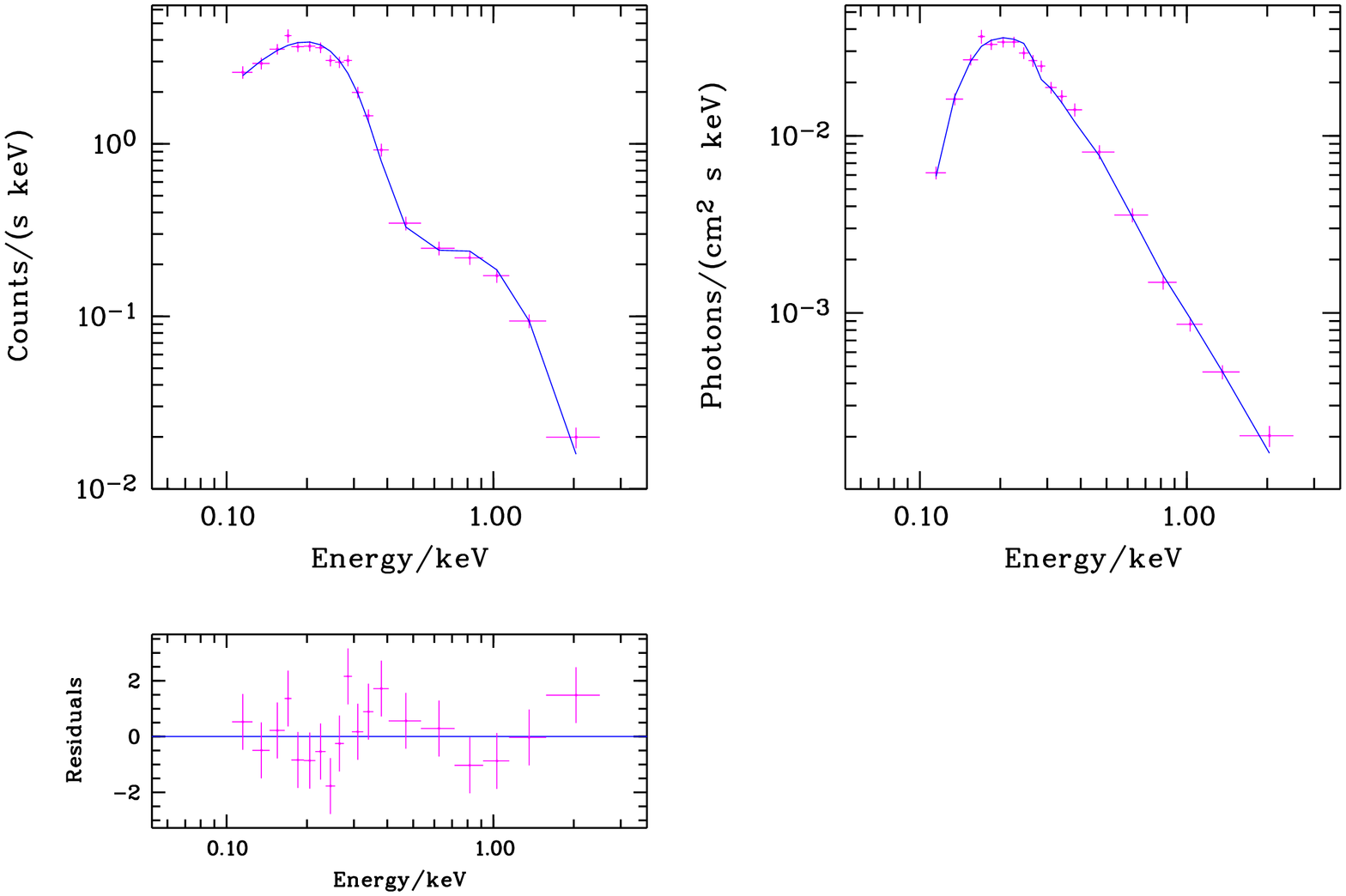,angle=-90,width=9cm,clip=}
\end{minipage}
\hfill
\begin{minipage}{9cm}
\psfig{figure=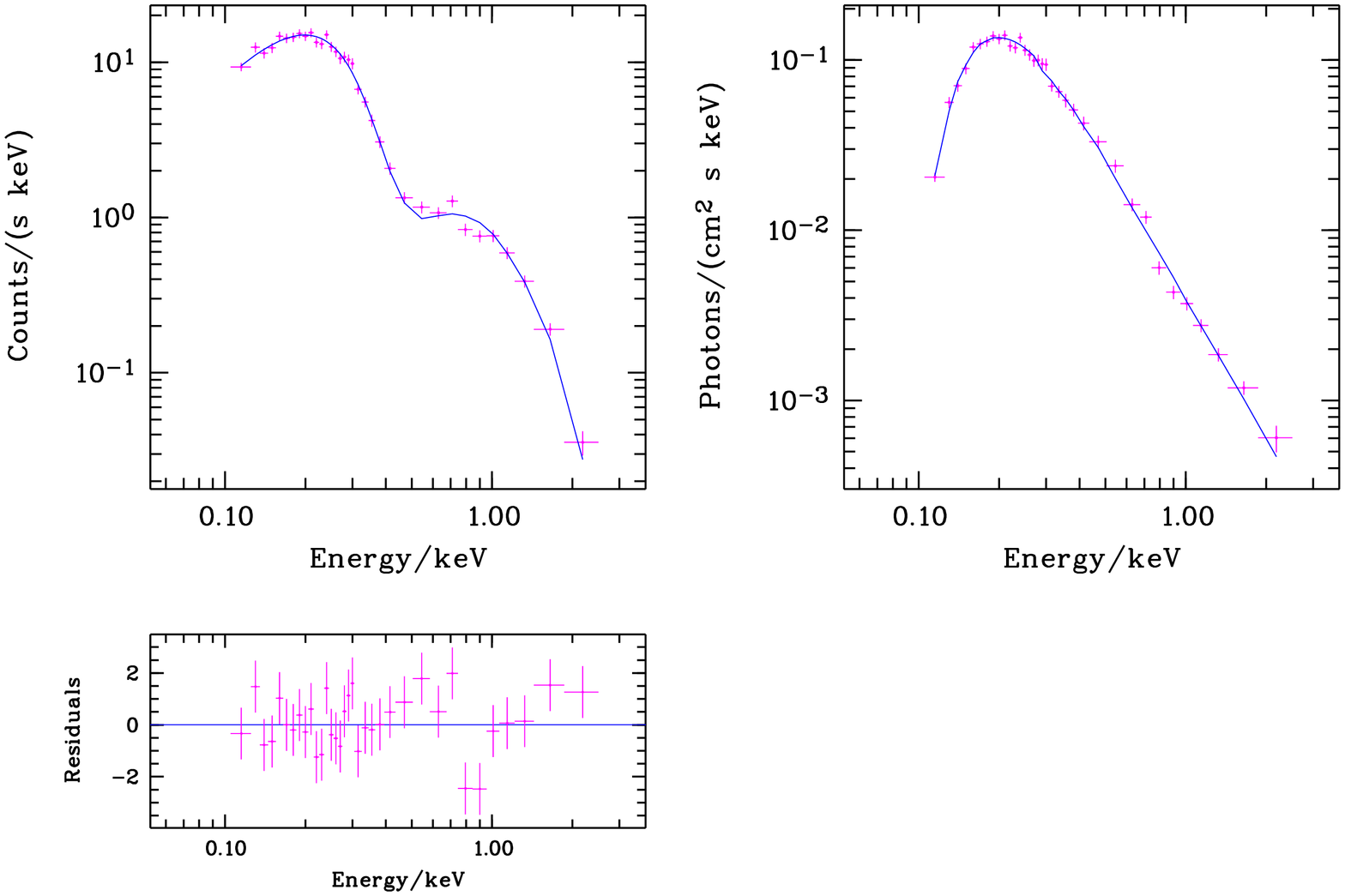,width=9cm,angle=-90,clip=}
\psfig{figure=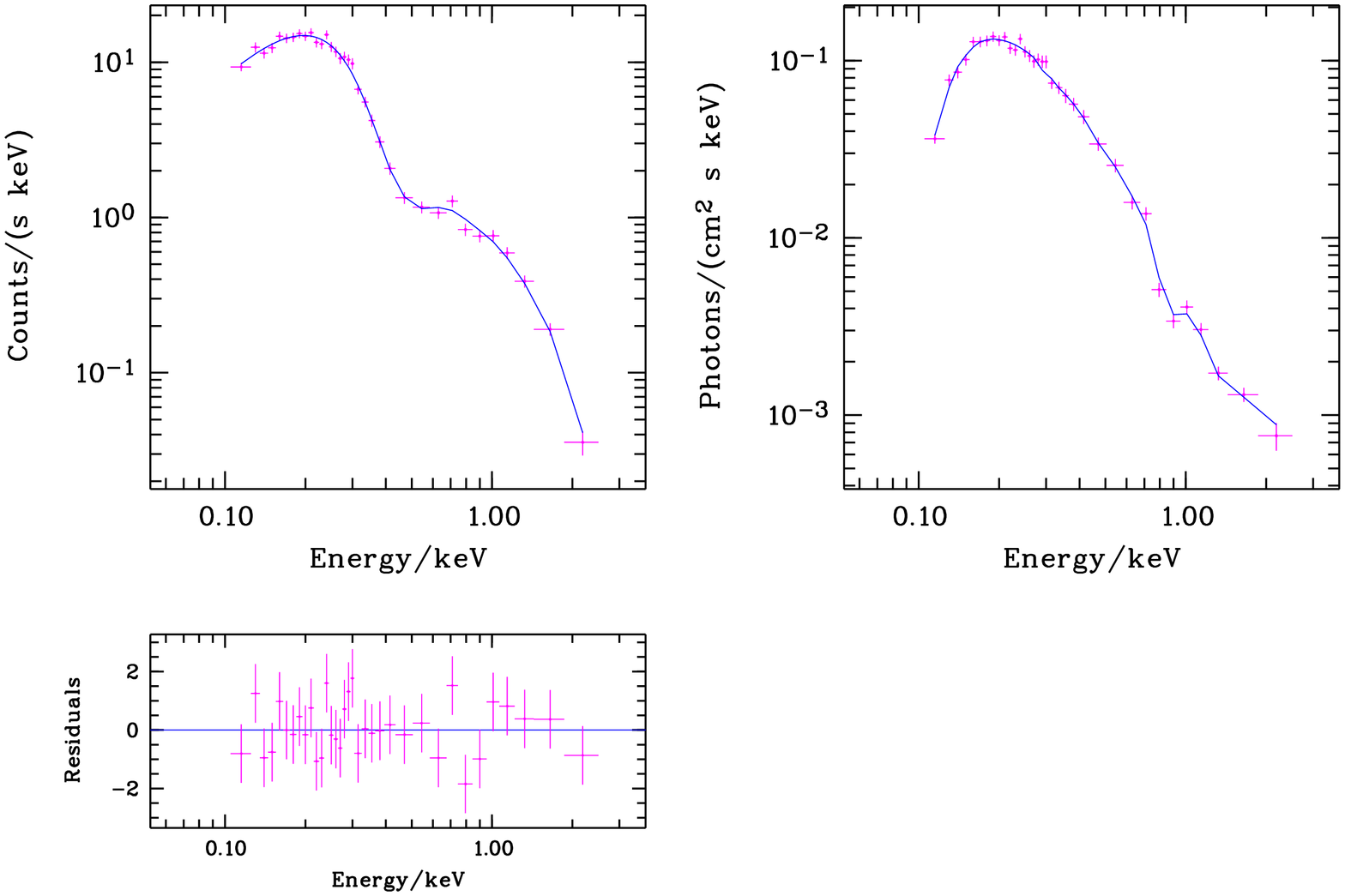,width=9cm,angle=-90,clip=}
\psfig{figure=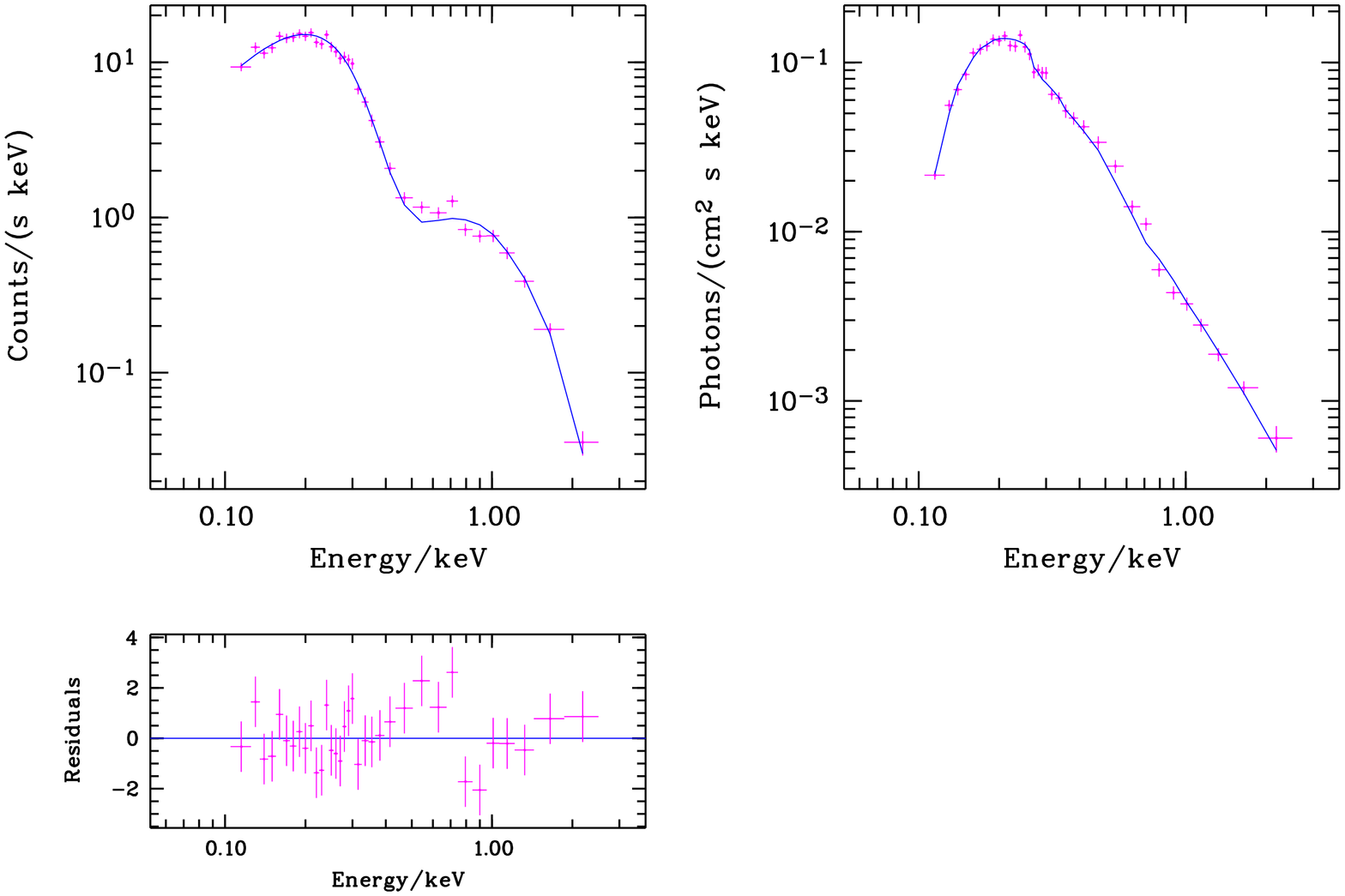,width=9cm,angle=-90,clip=}
\end{minipage}
\caption{The residuals of a simple power law fit (top), a warm absorber (middle)
and a {\it dusty} warm absorber fit to both PSPC observations. Left: P1,
right: P2.}
\label{residuals}
\end{figure*}

To illustrate the influence of dust on a warm absorber, we show in Fig.~\ref{wa-dwa} the 
model spectra for a dust-free and a dusty warm absorber model in the ROSAT
PSPC energy range. In the dust-free case the observed spectrum, when it is described with
a simple power law model, appears to be steeper than the intrinsic power law continuum 
because of a sequence of absorption edges due to ionized oxygen and 
neon. If dust is added to the warm absorber, the observed spectrum changes dramatically.
A carbon absorption edge appears at 0.28 keV due to the graphite species of dust and the 
absorption edges of ionized oxygen are less pronounced, because of the dust-induced lower 
ionization state of the warm absorber gas (Komossa \& Bade \cite{komossa98}). In particular, 
the \oxa\ edge is considerably weaker relative to \oxs\ (and almost completely disappears 
in the case of \iras). Contrary to the dust-free warm absorber, 
the presence of dust leads to an {\it apparent} flattening of the X-ray spectrum in
the ROSAT PSPC energy band.

\subsection{Results}

\subsubsection{Pointed Observations}

First, we applied various `standard models' to the data, e.g. a simple power law and
a power law plus black body, to check whether the results of BFP96 could be reproduced.
The residuals of the power law fit to the 
PSPC data are displayed in the top panels of Fig.~\ref{residuals}. We obtain $\Gamma = 
2.80\pm 0.15$, $N_{\rm H} = 1.10^{+0.28}_{-0.25}$ cm$^{-2}$ and $\chi^2_{\rm red} = 1.55$ 
for P1 and $\Gamma = 2.83\pm 0.11$, $N_{\rm H} = (1.22\pm 0.20)\times 10^{20}$ cm$^{-2}$ 
and $\chi^2_{\rm red} = 1.72$ for P2. Clearly, simple power law 
models are ruled out for both observations. 

In a second step, a {\em dust-free} warm absorber model was applied. In contrast to the 
approach of BFP96 we use the additional information on the hard X-ray power law available 
from the ASCA observation, i.e. we fix the intrinsic power law index to $\Gamma_{\rm ASCA} 
\simeq 2.2$ (Brinkmann et al. \cite{wpb96}; see also Sect. 4.2). We find an 
acceptable fit of the X-ray spectrum in terms of a dust-free warm absorber, with log 
$N_{\rm w} = 22.5\pm0.3$ ($\chi^2_{\rm red}= 1.05$) and log $N_{\rm w} =22.7\pm0.2$ 
($\chi^2_{\rm red} = 0.84$) for P1 and P2, respectively (Fig.~\ref{residuals}, 
middle panels). The value of $N_{\rm w}$ is consistent with that of BFP96. The amount of 
cold absorption in this model is slightly below, but still consistent with the Galatic value 
(\nh = $(0.08\pm 0.02)\times 10^{21}$ cm$^{-2}$, as compared to $N_{\rm H,gal} = 
0.11\times 10^{21}$ cm$^{-2}$). 

We conclude that the {\em dust-free} warm absorber model provides a successful fit to the 
ROSAT X-ray spectra and removes almost all systematic residuals. Moreover, the intrinsic 
X-ray continuum is consistent with the hard X-ray photon index obtained with ASCA. 

If we now modify this best fit model by adding the appropriate amount of dust to the warm 
absorber which corresponds to the value of $N_{\rm w}$, taking into account the gas-dust and 
gas-radiation interaction, the expected X-ray spectrum changes drastically and the data 
cannot be fit at all ($\chi^2_{\rm red} \sim 150$). This is partly due to dust-induced 
additional absorption, in particular at the carbon edge, which leads to a much flatter spectrum 
than observed. This still holds if we allow for non-standard dust, i.e., if we selectively 
exclude either the graphite or silicate species. A more detailed exploration of the very
large paramater space of dust properties is not warranted with current data and should await
the availability of higher resolution X-ray spectra.    

The expected column density derived from optical extinction is much smaller 
than the value of $N_{\rm w}$ determined from the dust-free warm absorber fit. Therefore, 
in a next step, we allowed all parameters (except $\Gamma = \Gamma_{\rm ASCA}$) to vary 
and checked whether a dusty warm absorber can be successfully fit at all. This is not the 
case. For example, if log $N_{\rm w}$ is fixed to log $N_{\rm opt}= 21.2$ we get 
$\chi^2_{\rm red}\sim 40$.

Next, the underlying power law spectrum was allowed to vary as well. Fixing $N_{\rm w}$ to 
$N_{\rm opt}$ again, the fit of the P2 observation results in a steep intrinsic spectrum 
with $\Gamma = 2.86\pm 0.04$ in order to compensate for the 'flattening effect' of dust 
(cf. Komossa \& Fink 1997a,b). The intrinsic continuum is much steeper than obtained 
with ASCA and still 
the fit is statistically not acceptable ($\chi^2_{\rm red} = 1.68$). The ionization parameter
is not well constrained by the data ($\log U = 0.19^{+0.84}_{-0.89}$). Similar results
are obtained for the P1 observation ($\chi^2_{\rm red} = 1.35$). The residuals 
between 0.6 and 1 keV remain (Fig.~\ref{residuals}, bottom panels) and a 
dust-free warm absorber model provides a better description of the PSPC data. 

\subsubsection{Survey observation}

A simple power law model provides an excellent fit to the Survey data with 
$\Gamma = -3.2\pm0.3$ and \nh = $(0.15\pm0.06)\times 10^{21}$ cm$^{-2}$ 
($\chi^2_{\rm red}$ = 0.74).

When fitting the warm absorber models to the Survey observation we get
the following results: A dust-free warm absorber with $\Gamma = -2.2$ provides a 
successful fit and the resulting model parameters are consistent with those of 
the pointed observations within the errors. If we add dust to the model and fix 
$N_{\rm w}$ to $N_{\rm opt}$, no successful fit can be achieved. For example, fixing 
\nh  to the Galactic value, the fit still yields an unacceptably high $\chi^2_{\rm red}$ 
of about 11. Allowing for a steeper intrinsic spectrum, with $\Gamma = 2.9$ as in the 
pointed observations, a dusty warm 
absorber with $N_{\rm w}$ = $N_{\rm opt}$ is consistent with the spectrum from the
Survey observation. However, the quality of the fit is slightly worse compared
to the dust-free warm absorber. 

We conclude that, within the limits of low photon statistics, the Survey data are roughly 
consistent with the pointed observations.

\section{The ASCA data}

The ASCA observation of \iras\ was analyzed and discussed in detail
by Brinkmann et al. (\cite{wpb96}) and Brandt et al. (\cite{brandt97}).
Therefore we describe in the following only the differences to the 
above papers in terms of data preparation and then turn to our new 
spectral analysis.
   
\subsection{Data preparation}

We used the 'Revision 2' data obtained from the ASCA public archive at Goddard
Space Flight Center (GSFC) and applied the following conservative screening 
criteria: For the GIS the minimum elevation angle above the Earth's limb ({\tt ELV}) 
was chosen to be $5\degr$. In the case of the SIS we used {\tt ELV}$ >10\degr$. 
To avoid atmospheric contamination, data were only accepted in the SIS when the 
angle between the target and the bright earth ({\tt BR\_EARTH}) was greater than
$20\degr$. A minimum cut-off rigidity ({\tt COR}) of 6 GeV/c was applied 
for both, the SIS and the GIS. Data taken within four read-out cycles of the CCD 
detectors after the passage of the South Atlantic Anomaly (SAA) and the day-night 
terminator are not considered in the analysis. Further, 
periods of high background were manually excluded from the data by checking the 
light curve of the observation. 

Source counts were extracted from a circular region centered on the target with a 
radius of $6\arcmin$ for the GIS and $4\arcmin$ for the SIS. We used the local 
background determined from the observation in the analysis for both detectors. In 
particular, the GIS background was estimated from a source free region at the same 
off-axis angle as the source and with the same size as the source extraction region. 

Both Brinkmann et al. (\cite{wpb96}) and Brandt et al. (\cite{brandt97}) neither find 
spectral differences between the two SIS detectors nor between the two observations
of \iras\ separated by about three days. We therefore merged the data from the two
observations to increase the signal-to-noise ratio. The same procedure was applied to 
the GIS detectors.

All spectra were rebinned to have at least 20 photons in each energy channel.
This allows the use of the $\chi^2$ technique to obtain the best fit values
for the model spectra. We used the latest GIS redistribution matrices
available (V4\_0) from the calibration database and created the SIS response
matrices for our observation using {\footnotesize SISRMG}, which applies the
latest charge transfer inefficiency (CTI) table ({\tt sisph2pi\_110397.fits}).
The ancillary response files for all four detectors were generated using the
{\footnotesize ASCAARF} program.

Spectra were fitted in the energy range 0.8 to 9 keV for the GIS and between 0.6 and
8 keV for the SIS. The upper energy boundaries are given by the maximum energy at 
which the source was detected in each instrument. The lower energy boundaries result 
from the calibration uncertainties of the detectors (e.g. Dotani et al. \cite{dotani}). 
If not mentioned otherwise, we quote the results of simultaneous fits to all four
instruments only.

\subsection{Spectral analysis}

\begin{figure}
\psfig{figure=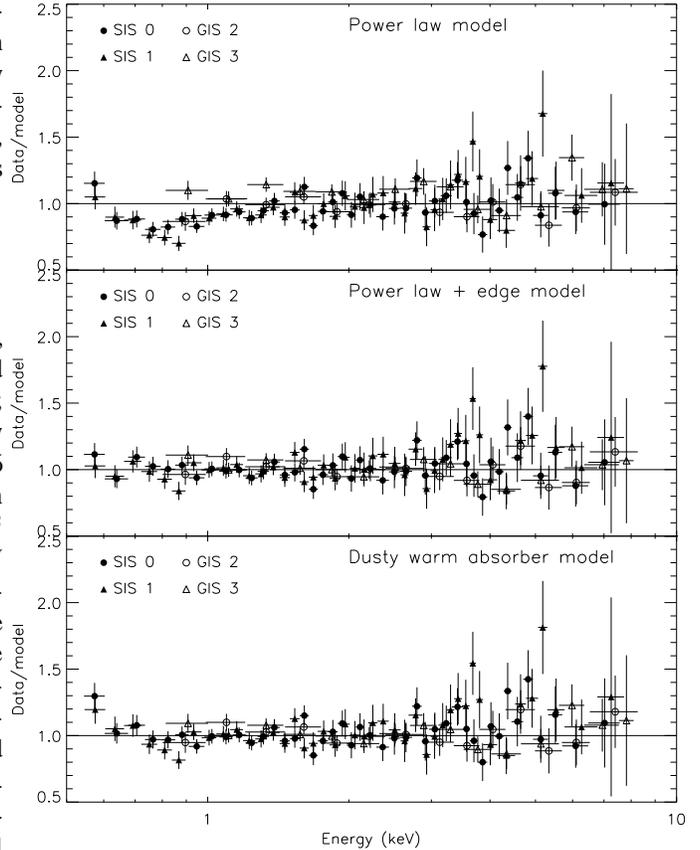,width=9cm}
\caption{The ratio of the ASCA data to various spectral models as described in the text. 
The four detectors are indicated by different symbols. The data have been rebinned according to 
the spectral resolution of the individual detectors.}
\label{ascafit}
\end{figure}

To check for the consistency of our data preparation procedure, we first fitted several
simple models to the ASCA data and compared them to previous analyses (Brinkmann et al.
\cite{wpb96}; Brandt et al. \cite{brandt97}; Leighly \cite{leighly}). 
For the simple power law model we get $N_{\rm H} = (1.75_{-1.67}^{+1.73})\times
10^{20}$ cm$^{-2}$, $\Gamma = 2.22\pm0.05$ and $\chi^2 = 548.3$ (565 d.o.f). The fitted 
\nh is consistent with the Galactic value and is therefore fixed to $1.1\times10^{20}$ 
cm$^{-2}$ in the following. Restricting the fit to energies above 2 keV we find $\Gamma 
= 2.24^{+0.08}_{-0.09}$ and $\chi^2 = 187.8$ (237 d.o.f.). In accord with Brinkmann et al. 
(\cite{wpb96}) and Leighly (1999), we note that the inclusion of a narrow ($\sigma = 0.05$ keV) 
iron K$\alpha$ line improves the fit significantly ($\Delta\chi^2 = -7.2$; P = 99.9\% according 
to an F-test for two additional parameters). The best-fitting parameters are: $\Gamma = 2.28\pm0.09$, 
$E_{\rm line} = 6.81^{+0.12}_{-0.11}$ and $EW = 313^{+107}_{-204}$ eV. The obtained iron
line parameters are typical for narrow-line Seyfert 1 galaxies (Leighly 1999). A gaussian 
emission line with the above parameters is included in all following models. Dividing 
the ASCA data by an extrapolation of this model to lower energies gives the residuals 
shown in the upper panel of Fig.~\ref{ascafit}. Clearly, there are systematic residuals 
below 1.5 keV. 

Adding a single absorption edge to the power law model results in the following best-fitting 
parameters: $\Gamma = 2.33\pm0.04$, $E_{\rm edge} = 0.75\pm0.04$, $\tau_{\rm edge} 
= 0.43^{+0.15}_{-0.12}$ and $\chi^2 = 509.5$ (563 d.o.f.). The agreement with the Brandt et al.
(\cite{brandt97}) results is excellent. As can be seen in the middle panel of Fig.~\ref{ascafit},
the fit is very good and all systematic residuals have disappeared.

As shown by Brandt et al. (\cite{brandt97}), a dust free warm absorber provides
a good fit to the ASCA data with $\Gamma = 2.31^{+0.07}_{-0.15}$ and 
log $N_{\rm w} = 21.59^{+0.15}_{-0.26}$.

Next we applied the dusty warm absorber model described in Sect. 3.2 to the ASCA data 
(including the iron K$\alpha$ line with line energy and $\sigma$ fixed to the above values).
The best-fitting model parameters are as follows: $\Gamma = 2.39^{+0.08}_{-0.07}$, log U = 
0.00$^{+0.31}_{-0.44}$, log $N_{\rm w} = 21.33^{+0.19}_{-0.26}$ and $\chi^2 = 524.4$ (563 
d.o.f.). The ratio of data to model for this fit is shown in the lower panel of 
Fig.~\ref{ascafit}. Note that the column density of the ionized material corresponds 
very well to the value expected from the observed optical extinction. Although the fit is 
statistically acceptable and clearly better than the simple power law model, it is 
significantly worse than the power law plus absorption edge fit. The data lie systematically 
above the model in the lowest energy channels. Whether this indicates that the dusty warm 
absorber model is not an optimal description of the ASCA spectrum or whether calibration 
uncertainties of the SIS detectors are important, is difficult to answer. 

\section{Discussion}

\subsection{A dust-free warm absorber?}

As we have shown in Sect.~3, a {\em dust-free} warm absorber model represents the 
formally best description of the PSPC spectrum of \iras. The systematic structures in the 
residuals almost completely disappear and, in addition, the spectral shape of the intrinsic 
X-ray continuum is consistent with the hard X-ray spectrum as observed with ASCA 
($\Gamma\approx 2.3$). We further note that applying the dust-free warm absorber model 
we also obtain a viable description of the ASCA data, although with significantly different 
physical parameters for the warm absorber as compared to the ROSAT PSPC results. The
energy of the absorption edge is significantly lower in the ASCA data and consistent with
\oxs, whereas the PSPC data clearly favor \oxa. Furthermore, the optical depth of the
absorption edge is much higher in the PSPC data and, consequently, the column density of 
the warm absorber derived from the ROSAT observations (log~$N_{\rm w} = 22.7\pm 0.2$) is 
about an order of magnitude higher than in the ASCA observation (log~$N_{\rm w} = 
21.59^{+0.15}_{-0.26}$; Brandt et al. 1997). 

The rapid X-ray variability observed in \iras\ indicates changes in the ionizing 
continuum and thus a variable ionization state of the warm absorber is not unlikely.
This might explain the change in the absorption edge energy as \iras\ was about
a factor of four brighter in X-rays during the second ROSAT observation (P2) as compared 
to the ASCA observation.  In the bright state, most of the oxygen is ionized to
\oxa. When the ionizing flux decreases, the \oxa\ ions recombine to \oxs\ and
the observed absorption edge energy decreases. This kind of variation of the ionization
state of the warm absorber has also been claimed for the Seyfert 1 galaxy MCG--06--30--15 
(Reynolds et al. \cite{reynolds}; Otani et al. \cite{otani}). The first PSPC observation 
(P1), when \iras\ was in a state comparable to the ASCA observation, suffers from low photon 
statistics and although the resulting spectral parameters of the dust-free warm absorber 
model are consistent with P2 we note that the higher $\chi^2_{\rm red}$ (1.05 compared to 
0.84) might indicate spectral changes.  

Apart from changes in the ionization state, the differences between the ASCA and ROSAT
spectra also require a change in the column density of the ionized material of at least
$4\times 10^{22}$ cm$^{-2}$ between the two observations, i.e. within about 3 years.
One might speculate that isolated clouds of ionized material stripped off the torus 
and moving across the line of sight could be responsible for such a change in column 
density. 

A principal difficulty of any dust-free warm absorber model of course is to explain the
discrepancy between the observed amount of optical reddening and the absence of any cold 
X-ray absorption. Apart from postulating atypical gas-to-dust ratios in \iras\ we 
might think of two possible ways out: variable optical extinction and different paths 
for the optical and the X-ray radiation.

With regard to the first possibility we note that the X-ray and the optical observations 
were not simultaneous and therefore variable optical extinction cannot be excluded. In fact, 
changes in the reddening of emission lines in Seyfert 1.8 galaxies on a time scale of years 
were reported by Goodrich (\cite{goodrich89a}, \cite{goodrich90}) and interpreted in terms 
of moving {\em cold} obscuring material. On the other hand, no changes in the optical 
extinction have been observed up to now (Wills et al. \cite{wills}; Lanzetta et 
al. \cite{lanzetta}) and the preliminary results of recent spectroscopic observations of \iras\ 
(Leighly, priv. com.; Papadakis, priv. com.) are completely consistent with the original Wills 
et al. (\cite{wills}) results. Furthermore, the amount of cold absorption measured with the 
ROSAT PSPC and ASCA 
is always consistent with the Galactic value and does not indicate variable cold absorption. 
Both results favor constant optical extinction. Obviously, {\em simultaneous} soft X-ray and 
optical observations of \iras\ are required to ultimately decide whether variable optical 
extinction plays a role, but based on current observations this seems highly unlikely.

An alternative explanation for the apparent discrepancy between optical and cold
X-ray absorption would be a special geometry, in which the optical and the 
X-ray continuum do not travel along the same paths to the observer. Since the observed
rapid X-ray variability of \iras\ effectively excludes a significant contribution of 
scattered X-rays, one might invoke a partial covering geometry, where the optical
continuum happens to pass through a lower total column density of obscuring material.
Although such a scenario might be constructed for a single object, this approach
becomes increasingly contrived in the light of the growing number objects, for which a 
'simple' dusty warm absorber provides a consistent explanation of the X-ray and optical 
properties (e.g., NGC 3227, NGC 3786, IRAS 17020+4544). We therefore now turn to the 
dusty warm absorber hypothesis.

\subsection{A {\em dusty} warm absorber?}

The prime argument for the presence of dust within the warm absorber is the
apparent discrepancy between the observed optical extinction and the absence of
cold absorption in the X-ray spectrum. Furthermore, as we have shown in Sect.4.2, a 
self-consistent Galactic-ISM dusty warm absorber model is a reasonable representation of the
ASCA spectrum of \iras\ and the column density of the warm absorber determined by 
spectral fitting is consistent with the \nh derived from optical reddening.

We note, however, that the ASCA spectrum is not very sensitive to the dustiness of 
the warm absorber, since the most important features (like the Carbon absorption edge)
are expected below 0.6 keV. Nevertheless, the absence of a strong \oxa\ edge in the ASCA 
spectrum\footnote{This is confirmed by Brandt et al. (\cite{brandt97}), who find $\tau = 
0.09^{+0.11}_{-0.09}$ for the \oxa\ absorption edge, which is consistent with zero.} 
(Brinkmann et al. \cite{wpb96}) is exactly the expected signature of dust within a warm 
absorber (cf. Fig.~\ref{wa-dwa}). The structure in the spectrum around 0.65 keV was modeled 
as an \oxs\ absorption edge by Brandt et al. (\cite{brandt97}) and as an isolated emission
line by Brinkmann et al. (\cite{wpb96}). In fact, even both features may be present and 
overlapping. We note that also an instrumental emission feature has been reported around 
0.6 keV  (see Sect.\,2.1 of Otani et al. \cite{otani}).  

The major problem with dusty warm absorber models for \iras\ lies in the fact that 
they are, at most, marginally consistent with the ROSAT PSPC data. Even if one accepts
the reduced quality of the fit ($\chi^2_{\rm red} \approx 1.7$), a temporal change of 
the intrinsic continuum of $\Delta\Gamma \approx 0.5-0.7$ between the ROSAT and the ASCA 
observation has to be explained. 

The difference in the quality of the fit between the dust-free and the dusty 
warm absorbers is due to the clear indication of an \oxa\ absorption edge in 
the ROSAT data (see also Table 1 of BFP96), which should be much weaker
if the warm gas is dusty. In principle, it is possible to introduce increased 
\oxa\ absorption also in the case of a dusty warm absorber, for example by 
increasing the gas-phase oxygen abundance (which is depleted in the models to account 
for the binding of metals into dust). However, the limited spectral resolution 
of ROSAT certainly does not warrant this kind of fine-tuning to achieve optimal 
fits and the improved spectral capabilities of future X-ray missions are be needed 
to study in detail the elemental abundances of the warm absorber as well as the 
properties of the any dust in \iras. 

As far as the change in the intrinsic continuum is concerned, one might speculate
about various explanations, including a 'real' change in the spectrum and instrumental
effects.

The X-ray continuum spectrum might indeed have varied by $\Delta\Gamma\approx 0.5-0.7$ between 
the ROSAT and the ASCA observations, which are separated by about three years. This possibility 
of course cannot be ruled out and in fact spectral changes have previously been observed in 
narrow-line Seyfert 1 galaxies (e.g. Mrk 766; Leighly et al. 1996). However, we consider 
it unlikely for \iras\ in view of the fact that no spectral variations have been observed 
between as well as within the individual PSPC and ASCA observations. In particular, the 
spectra observed in the two PSPC observations seem to be consistent despite a dramatic
increase in flux by a factor of four. Furthermore, no spectral variations have been
detected within the ASCA observation (e.g. Brinkmann et al. \cite{wpb96}). 

Systematic differences between ASCA and ROSAT spectra have been reported for various simultaneous 
observations in the past (e.g. NGC 5548; Iwasawa et al. (\cite{iwasawa}) and references therein). 
In particular, the photon index tends to be steeper by $\Delta\Gamma\approx 0.4$ in these PSPC 
observations as compared to ASCA. On the other hand there are also observations, where
the ASCA and ROSAT spectra do agree reasonably well (e.g. Miyaji et al. \cite{miyaji}; Brinkmann
et al. \cite{brinkmann98}; Cappi et al. \cite{cappi}). Although these observations were not done 
simultaneously, it seems unlikely
that spectral changes occured in all these sources which exactly compensate for any putative 
calibration uncertainties. The ASCA/ROSAT discrepancies definitely need to be further 
investigated. This is, however, beyond the scope of this paper.

We conclude that it cannot be excluded that the observed difference between the ROSAT and ASCA 
power law continuum spectrum of \iras\ is at least partly due to instrumental effects. We note, 
however, that even if we allow for the steeper PSPC spectrum in the dusty warm absorber model 
(and only in this model a systematically steeper continuum is required for the ROSAT data), the 
fit is still only marginally acceptable.

In the previous section we discussed the possibility of a variable dust-free warm absorber. 
In order to preserve the dusty warm absorber hypothesis, and hence to obtain a 
physical model to explain all X-ray {\it and} the optical properties consistently, we 
might also speculate about more complicated warm absorber models. For example,
a two-zone warm absorber has been proposed for the Seyfert 1 galaxy MCG--6--30--15
(Otani et al. \cite{otani}), i.e. a variable inner warm absorber responsible for 
the \oxa\ absorption edge and an outer warm absorber in a lower ionization state 
and thus mainly imprinting the \oxs\ edge. \iras\ might be a similar case: the inner 
warm absorber varied between the ROSAT and the ASCA observation either by changes in 
the ionization state and/or by a variation in N$_{\rm w}$, whereas the outer {\it dusty} 
warm absorber is present in both observations. 

\section{Conclusions}

We presented new ROSAT HRI data for \iras\ and a re-analysis of the ROSAT PSPC and ASCA
spectra in terms of a warm absorber, but self-consistently including the effects of
dust on the soft X-ray spectrum. 

The HRI light curve of \iras\ confirms the rapid and large amplitude variability
already noted in the ASCA data (Brinkmann et al. \cite{wpb96}). We observed a factor 
of two change in count rate within one day and 20\% variability within about three
hours. These rapid variations of the soft X-ray flux rule out a significant 
contribution of scattered X-rays to the total flux. 

The analysis of the measured ROSAT PSPC and ASCA spectra gives contradictory results.
The PSPC data are best fit by a dust-free warm absorber, but large changes in the
column density of the ionized material and the ionization state are required to describe
the ASCA data with this model. And finally it cannot explain the discrepancy between 
the optical reddening and the absence of cold X-ray absorption. On the other hand, a dusty
warm absorber is consistent with the ASCA data and gives a column density of the warm
material, which agrees very well with the one predicted from optical reddening. However,
the dusty warm absorber is, at most, marginally consistent with the PSPC data and requires
a much steeper ($\Delta\Gamma\approx 0.5-0.7$) intrinsic continuum as that found in the
ASCA data.  
  
Physical effects, such as a variable warm absorber column density and ionization 
state as well as instrumental effects have to be considered to explain the 
apparent discrepancies. Maybe a two-zone warm absorber, similar to the one postulated
for the Seyfert 1 galaxy MCG--6--30--15, is able to provide a consistent explanation
of all X-ray data and the optical extinction, but we consider it an over-parameterization
of presently available X-ray data. From spectral fits to current ROSAT and ASCA data alone the 
presence of dust within the warm absorber cannot be proven and the discrepancy between 
the observed optical reddening and the absence of cold X-ray absorption remains the strongest
argument in favor of a {\it dusty} warm absorber in \iras. Future X-ray observations 
(with XMM or probably SAX), will be able to directly probe the dust content of the
warm absorber, e.g. via the predicted Carbon absorption edge.    

\section*{Acknowledgements}
The ROSAT project is supported by the Bundesministerium f\"ur Bildung, Wissenschaft,
Forschung und Technologie (BMBF) and the Max-Planck-Gesellschaft. 
This research has made use of the ASCA IDL Analysis System developed by Tahir Yaqoob, 
Peter Serlemitsos and Andy Ptak and of the NASA/IPAC Extragalactic Data Base
(NED) which is operated by the Jet Propulsion Laboratory, California
Institute of Technology, under contract with the National Aeronautics
and Space Administration. We thank Gary Ferland for providing {\sc Cloudy}.

\end{document}